\documentclass[10pt,journal,final,finalsubmission]{IEEEtran}
\usepackage{color}
\usepackage[usenames,dvipsnames,svgnames,table]{xcolor}
\usepackage{algorithm}
\usepackage[bf,SL,BF]{subfigure}
\usepackage{algorithmicx}
\usepackage{algpseudocode}
\usepackage{graphicx}
\usepackage{subfigure}
\usepackage{amsmath}
\usepackage{url}
\usepackage{mathabx}
\usepackage{sidecap}

\newcommand{\apply}{\Gamma}
\newcommand{\bind}{\beta}
\newcommand{\func}[1] {\underline {#1}}
\title{Pipelined Workflow in Hybrid MPI/Pthread runtime for External Memory Graph Construction}
\author{\IEEEauthorblockN{Sandeep Gupta}\\
\IEEEauthorblockA{San Diego Supercomputer Center\\
University Of California, San Diego\\
San Diego, California, 92093\\
Email: sandeep@sdsc.edu}
}
\begin{document}
\maketitle
\begin{abstract}
Graph construction from a given set of edges is a data-intensive operator 
that appears in social network analysis, ontology enabled databases, and, other 
analytics processing. The operator represents an edge list  to what is called compressed sparse row (CSR) representation 
(or sometimes in adjacency list, or as clustered B-Tree storage). In this work, we show how to scale CSR construction to massive scale
 on SSD-enabled supercomputers such as Gordon using pipelined processing.
We develop several abstraction and operations for external memory and parallel edge list and integer array processing that are 
utilized towards building a scalable algorithm for creating CSR representation. 

Our experiments demonstrate that this scheme is four to six times faster than  currently available implementation. Moreover,
our scheme can handle up to  8 billion edges (128GB) by using external memory as compared to  prior schemes where performance degrades considerably for edge list
size 26 million and beyond.

\end{abstract}

\section{Introduction}
Web click stream, online social networks, semantic web  etc. are some of the domains where inherently graph structured data is produced as a stream of edges or triples. Consider, for example, a set of tuples (product,customer) produced from a high traffic e-commerce website indicating the products viewed by its customers in last hour. Collection of pairs (source, destination) generated within a Twitter network that indicate message exchanges between its users is another scenario where edge list arises. Many large  data stores are maintained in Resource Description Framework (RDF) format which represents data as collection of edges. Such data stores are often very large. As an example DBpedia consisting of  approximately  2 billion RDF triples  returns a large collection of triples.

In all these scenarios the edge or triples data set is best converted into graph representation, i.e., as an adjacency list, compressed sparse row format (CSR), or clustered B-Tree indexed storage. This conversion allows for efficient graph analysis and navigation over the received data. For example, a market analyst would be interested in finding the most connected bipartite graph of product and customers. In Twitter dataset we may want to know the most influential twitterer (a person whose tweet diffused the most) in last one hour. An ``informatican'' (domain knowledge engineer) would like to perform guided navigation on the graph representation  of the returned RDF triples. Such a navigation can either be  performed using variations of Sparql such as nSparql, or pSparql~\cite{sparql}( and references therein) or through hard coded algorithms for finding graphs that are homeomorphic to a given pattern~\cite{sparql_graph_homomorphism,Neumann:2008:RRE:1453856.1453927}. 

\begin{figure*}[!ht]
\centering
  \includegraphics[width=6.50in]{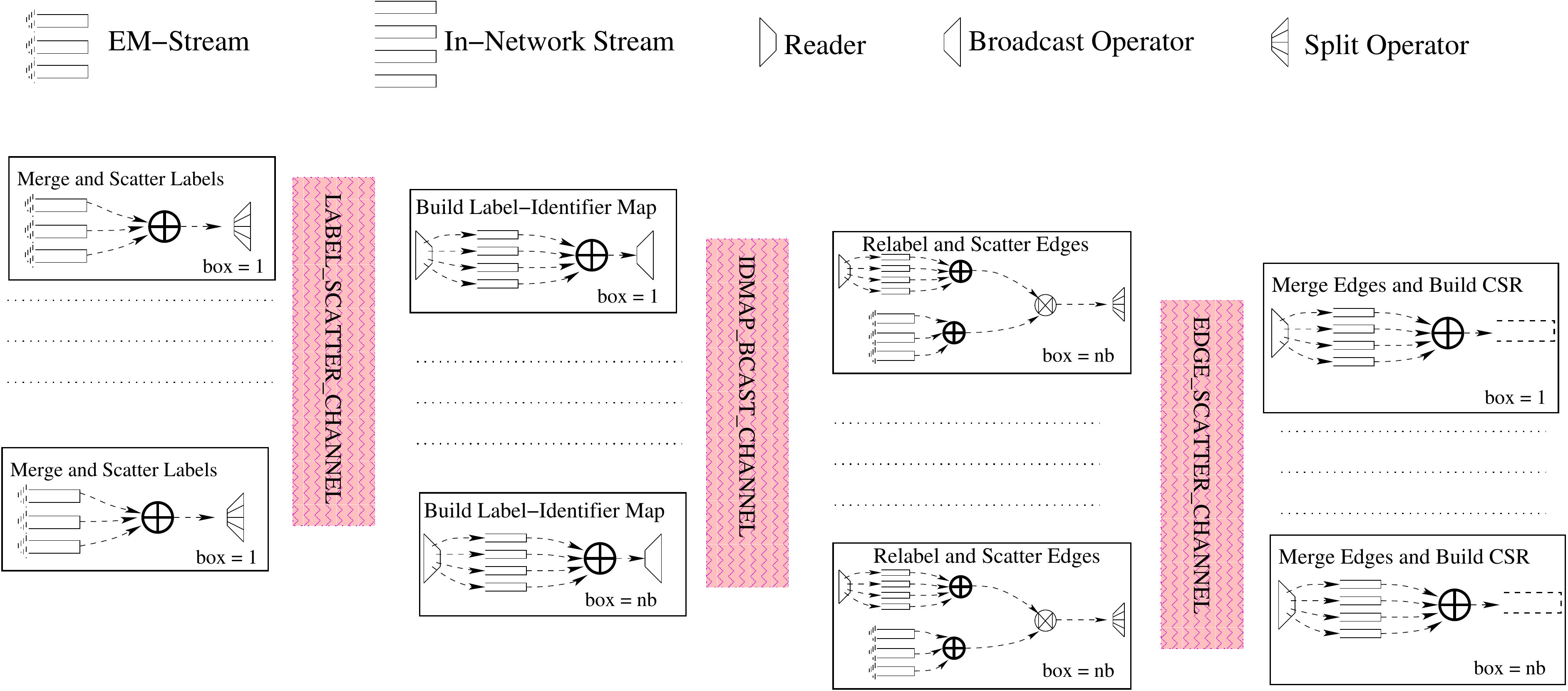}
  \caption{Processing for building CSR as a pipelined workflow in MPI/Pthread hybrid model of computation. \label{fig:pipeline}}
\end{figure*}

One of the most important use of conversion from edge list to graph representation arises in the world of Ontology-enabled databases\cite{Calvanese:2009:ODD:1611703.1611710,Dolby:2008:SGC:1483155.1483189,OntDB}. Such databases are popular  in medical and life-sciences domains. They have two distinct but connected parts: the ontology and annotation component and instance and experiment data component. The ontologies and annotations describe the metadata (and their relationships) while instance data consists of entities, experimental readings,  and their relationships.  A typical query in such databases  selects a subgraph from the metadata component. The returned sub-graph is the set of nodes from the instance domain which have links from the selected metadata subgraphs. Further processing on the returned sub-graph is a very common step.  For example, a user may select all ``Gaba-ergic neurons'' and its sub-type in the metadata component. 
The system then returns all the publications, experimental findings, biological entities  that are annotated/have connection from at least one of the metadata nodes. Following this, a user may request to find the most ``central'' publications or perform path pattern query on the returned node. Again, it is efficient to build the graph representation over the selected subgraph to answer such queries efficiently.

The problem of scalable algorithms for edge list to graph construction (either in external memory or main memory) has been studied less frequently in literature. Most RDF processing toolkits primarily provide sequential, main memory or relational table based storage and retrieval of RDF triples~\cite{Wood05kowari:a,DBLP:conf/semweb/McBride01,Rohloff:2007:ETT:1780453.1780500,triple/Store/Report}. To the best of our knowledge, we do not know of any parallel (shared memory) implementation of RDF-toolkit. Some of the RDF libraries provide techniques to build the graph on external memory (disk). In most work the edges are stored as a relational table.  A B+-tree index structure, primarily over the source field, is constructed as well. Among the set of existing tools, B+-tree is the most efficient method for external memory graph representation of a given edgelist. In~\cite{Weiss:2008:HSI:1453856.1453965} showed that B-Tree index does not work well for graph style workload. 

Customized storage mechanisms have also been proposed for RDF triples~\cite{Weiss:2008:HSI:1453856.1453965}. These mechanisms primarily focus on optimizing queries expressed using Sparql language. Such queries are conjunctive selection and join predicates and do not express navigation or higher order  graph operations such as BFS, centrality etc. 
Distributed RDF processing, especially using the map-reduce framework,  has been an active area of research~\cite{scalable-subgraphs,Quilitz:2008:QDR:1789394.1789443,Rohloff:2010:HMS:1940747.1940751,Stuckenschmidt:2005:TDP:1358613.1358616}. Here again the focus has been on expediting queries expressed under the restricted framework of Sparql. In addition to these, several distributed general purpose graph processing engines have been proposed recently~\cite{Low:2012:DGF:2212351.2212354,Malewicz:2010:PSL:1807167.1807184,Buluc:2011:CBD:2076556.2076566}. Although each of these scheme has some mechanism for graph ingestion, the focus is primarily on optimizations for processing of graph operators such as BFS, pagerank, centrality etc and not on edge list to graph construction.

The Parallel Boost Graph Library (PBGL)~\cite{gregor06:pbgl_siampp_presentation}  addresses edge list to CSR construction  in distributed settings. The library provides  several distributed mechanisms for constructing the CSR representation of graph from edgelist. However, with the exception of one, all the rest of proposed mechanisms require that edges be present on all compute nodes and require additional temporary main memory storage. A mechanism is also provided  that does not have the requirement and can handle edges being distributed across the compute-nodes. 
This scheme performs in-place construction of graph representation (CSR) if 
the edges are represented with two vectors S and D,0 i.e., pair  $(S[i], D[i])$ is an edge for any index $i \in [0:m]$. This scheme can handle much larger edge lists  than prior scheme but it induces too many memory swaps thereby exponentially increasing the processing time. 
A shared memory parallel graph library is presented in~\cite{Madduri11}. However, it does not address the problem of graph construction in parallel. Once the graph is constructed it provides routines for several graph operations such as BFS, connected component etc.

\begin{figure}[!ht]
  \includegraphics[width=2.50in,angle=-90]{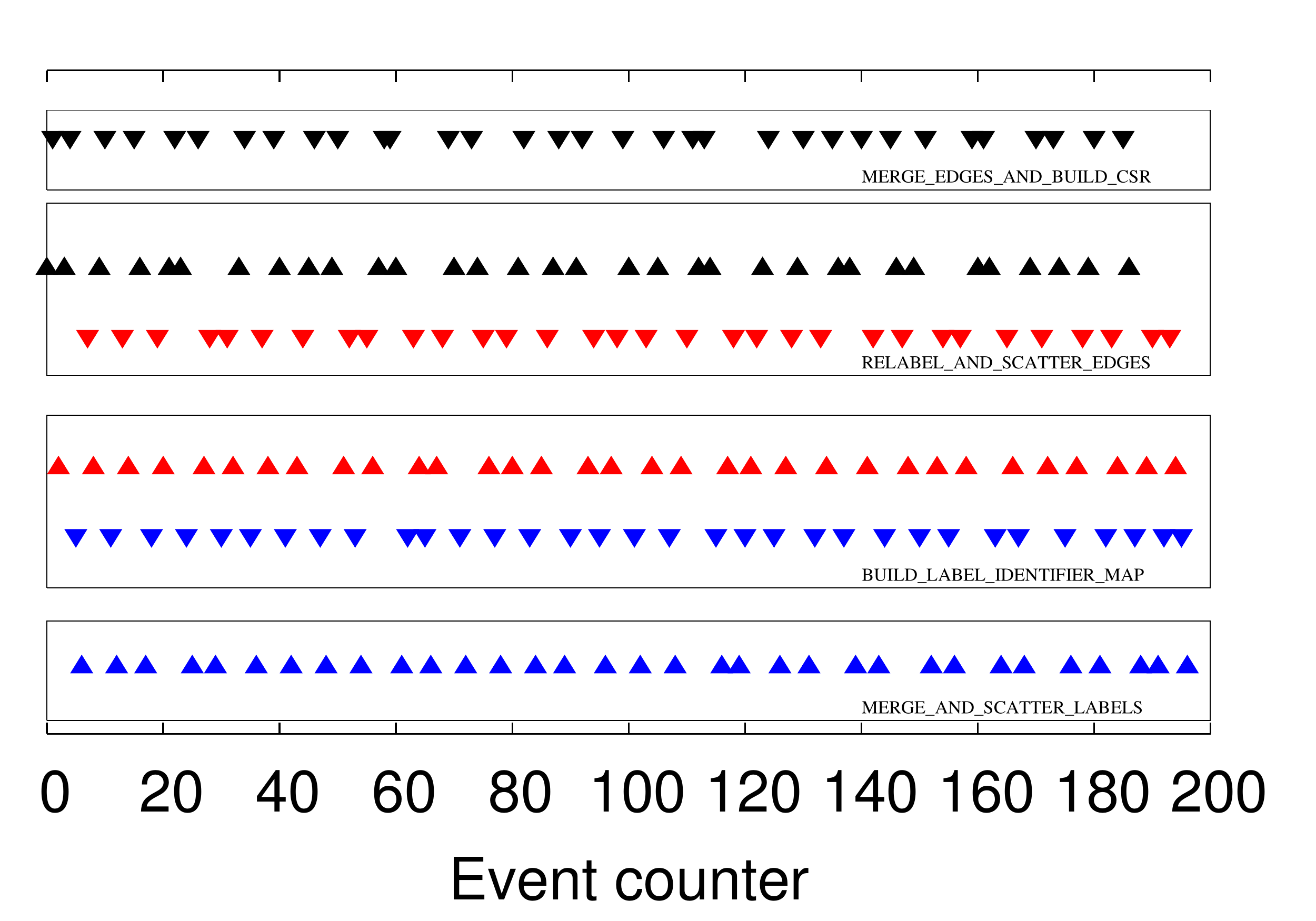}
  \caption{Illustration of pipelined processing.  Upper facing triangle represent a message send event while bottom facing triangle 
    represent message received event. The top row (black colored triangles) shows message exchanges over the EDGE\_SCATTER\_CHANNEL, middle row (red colored triangles) for IDMAP\_BCAST\_CHANNEL, and, the lowest 
    row (blue colored triangles) for LABEL\_SCATTER\_CHANNEL. The rectangular boxes represent the stages. A triangle within a box implies that the communication was initiated at the stage represented by the box. For example, all send (upper triangle) over IDMAP\_BCAST\_CHANNEL is performed at build label idmap stage.\label{fig:timestamp}}
\end{figure}

In this paper, we consider a hybrid distributed+shared memory out-of-core approach for building graph representation from a collection of edge lists.  Specifically, we consider the problem of building distributed graph representation of a given set of edges (or triples) in the MPI/Pthread  framework when the edge list and graph representations are maintained in external memory (preferably in SSDs).
To the best of our knowledge this variation of problem has not been studied before. On a single laptop we can process 32GB graph in less than 1400 secs and 128GB graph in 2400 secs.
The primary contribution of our approach is that it  alleviates the limitation on main memory and can scale to massively large number triples (2 billion edges) and beyond. With our scheme the only true limitations  are the number of cores available per CPU, the bandwidth and IOPS of the external memory, and, the communication cost associated with the interconnect fabric.
 
All  these aspects are growing very fast in the industry. Multi-core cpus with 32-128 cores and potentially 1000s of cores would be possible on modest enterprise class CPUs while the bandwidth and IOPS of modern  SSDs and phase change memory devices are upwards of 5600 MB/s and  1.2M IOPS (assuming 8K block size) respectively~\cite{OCZ}. The interconnects costs too has driven down significantly. These trends together with our distributed algorithm imply that in near future massive scale graph (64 billion edges)  processing would be feasible even on modest size computers (32--64 compute nodes).

\paragraph*{Contribution}
Our contribution in this paper are as follows:
\begin{itemize}
\item We propose a set of generic  abstractions and operators for data manipulation and movement over out-of-core (external, SSD) storage and network subsystem.  

\item We provide efficient implementation of these abstractions within our framework which consists of clusters of solid state enabled machines connected over high bandwidth low-latency network. 

\item We propose a buffered network reader, a novel scheme which allows us to pipeline operators in the hybrid MPI/Pthread communication framework without any additional synchronization overhead. Pipelining operators is crucial because it avoids the need for reading and writing massive amounts of data in external memory. An additional benefit of pipelining is that it can produce results early (and possibly in sorted order) which in same cases has been shown to aid further optimization. 

\item We design a scheme for building distributed CSR representation. The scheme exploits pipelining, SSDs, mulit-cores, and, fast interconnect to scale this operation to billions of edges. Figure~\ref{fig:pipeline} shows the pipeline stages and the communication channels between them for constructing distributed CSR representation. Figure~\ref{fig:timestamp} shows the flow of messages across the stages in the pipeline. We see a workflow with smoothed movement of data. 

\item Our experiments show that the scheme is 4 --6 times faster than standard implementation and that it can handle 10 time bigger size edge list. Our experiments identify limitations with the current hybrid MPI/Pthread runtime that limit the scalability to two compute nodes.

\end{itemize}

The rest of the paper is organized as follows: We begin with definitions and then formally setup the problem and our objective. We present all the abstractions, operators, and their implementation in the hybrid MPI/Pthread framework. We then describe our procedure for building distributed CSR representation using these operators. 
Finally, we present experiments  that study the performance trade-offs and scalability of the proposed approach.

\subsection{Setup and Problem Definition}
A cluster of $nb$ compute nodes, each with $nc$ cores, are connected via high bandwidth network, possibly an interconnect. 
A few  notations used to setup the problem of building CSR representation is as follows:
\begin{itemize}
\item Label: A character string used as an identifier to an object.
\item Vertex: A node in the graph and represented as $u, v,w$ etc. The vertex is either identified through a label or a numeric identifier.
\item Edge:  A pair $(u,v)$.
\item Adj(u):  List of vertices adjacent to $u.$
\item Graph: G= (V,E) represents a graph with $n = |V|$ vertices 
and $m = |E|$ edges.
\item $Induced Edgelist (V'):$ Given a subset $V' \subset V$, its  induced edge list is the subset of edges $E'$ whose source belong to $V'.$

\item $CSR(G)$: Compressed sparse row representation (CSR) of any graph $G$ consists of two vectors, namely, an offset vector ($offv$) and an adjacency vector ($adjv$). 
The $offv$ indexes into the $adjv$ vector. The adjacency (edge) information is stored in  $adjv$ vector. The neighbors of node $nid$ are stored as entries in $adjv$ vector from  range $[offv[nid], offv[nid+1]].$

\item $box:$ A compute node in the cluster is termed as a box.
\item $nb:$ Number of compute nodes (boxes) in the cluster. 
Symbols $box_0, box_1, \dots, box_{nb-1}$ identify 
box numbered as $0,1,\dots$, and $nb-1$ respectively.
\end{itemize}

\paragraph*{Problem Definition}
We are given a  collection of pair of labels (representing edge list) $E_l = \lbrace e_1=(u_1, v_1), e_2 = (u_2, v_2), e_3 = (u_3, v_3) \dots \rbrace$.
A label in this setup uniquely identifies a vertex and $E_l$ represents an edge list that induces a graph. 
Our objective is to  find a compressed sparse row representation of $E_l$, i.e., first, each node $u$ is assigned a unique numeric identifier 
$nid$ from range $[0:t]$, where $t$ is the total number of unique labels. We call this operation as {\em relabeling}. Second, build two arrays, the offset vector $offv$ and 
adjacency vector $adjv$, such that for any label (or vertex) $u$, its adjacency entries are in $adjv$ from index $offv[nid]$ to $offv[nid+1]$, if $nid$ is the numeric identifier for node $u$. 

Furthermore, we require the CSR representation be distributed across the $nb$ boxes. In order to achieve this, each label is mapped to a unique box. Let $L_b$ be the set of labels at box $b$.  Each label is assigned a unique local identifier between $0$ to $|L_b|$. The global identifier of a label is a pair consisting of the box id the label is mapped to and its local identifier at that box. 

\begin{figure}
\includegraphics[width=3.00in, height=2.00in]{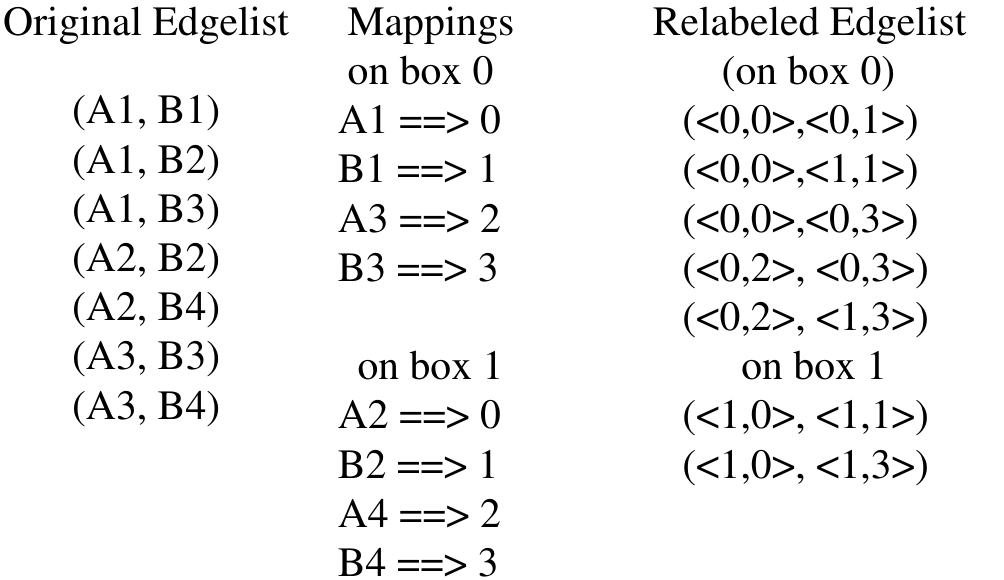}
\caption{Distribution and relabeling of a sample edge list over two boxes assuming all odd labels on box 0 and even labels on box 1. The relabeling assumes lexicographic ordering of labels per box. The right most columns depicts the relabeled edges on both the box. \label{fig:example}}
\end{figure}

The edges in $E_l$ are distributed  in accordance with the label partitioning, i.e., edges are placed on the same box on which its source is mapped. Let, $E_{l_b} = InducedEdgelist(L_b)$ denote the set of edges placed at box $b$. We also say that edges in $E_{l_b}$ are {\em owned} by box $b$. 
In the  distributed CSR representation for the edge list $E_l$, each box $b$ has an  $offv$ and $adjv$ vectors that define the CSR representation  $E_{l_b}$. The  entries in $adjv$ are the global identifier of the adjacent node. Figure~\ref{fig:example} shows relabeling and distribution performed for a sample edge list over two boxes assuming odd/even number based mapping from label to box. The local identifiers are assigned based up on the lexicographic ordering of the labels at the box. 
The final column shows the relabeled edges.

Finally, we observe that after relabeling, if the edge list $E_l$ (or $E_{l_b}$), after relabeling, is sorted based upon source field then building CSR representation is straightforward as shown in algorithm~\ref{alg:buildcsr}.
\begin{algorithm}
\caption{build\_csr(edge list el)\label{alg:buildcsr}}
\begin{algorithmic}[1]
\Require edge list $el$ be sorted 
\State aidx = 0, offv[0] = aidx, elidx = 0, csrc = 0
\While {1}
\If {$csrc \neq el[elidx].src$ }
\State ++csrc 
\State offv[csrc] = elidx
\State continue
\EndIf 
\State adjv[elidx] = el[elidx].des, ++elidx
\EndWhile
\end{algorithmic}
\end{algorithm}

Hence, we can reformulate the problem of building CSR representation as that of (a) distributing the labels, (b) assigning unique local identifiers to the labels (we term the mapping from labels to identifiers as {\em identifier map} (c) partitioning and distributing $E_l$ based upon the source field (d) relabeling destination field with the global numeric identifier e) relabeling source field local numeric identifier, and, finally (d) sorting $E_{l_{b}}$  at each box based upon the relabeled source identifier.

\section{Our approach for Building External Memory based Distributed CSR graph using hybrid MPI/Pthread parallel programming model}
We now describe our approach for building the distributed CSR representation assuming an unordered collection of edge list. We begin with few definitions followed by functions (utilities) needed from operating system/runtime, such as file I/O, mmap, communication primitives etc. We then define high order operators and primitive and provide implementation using the system provided calls and utilities.  Finally, the complete procedure for building distributed CSR is put together using these operators.

\paragraph{System Utilities and Functions}
The list of   systems calls and runtime utilities that are used by our algorithm for building distributed CSR:

\begin{itemize}
\item Storage cost $\mathcal S(\tau)$: Gives the number of bytes required to store a particular data object of type $\tau$. For example ${\mathcal S}(int)$ is 8 bytes in our setup. Similarly, an edge requires 16 bytes as it a pair.
\item $ptr \longleftarrow mmap(fn, offset, sz)$ : map the data in file  $fn$ starting from $offset$ and of $sz$ bytes in the virtual address space of the calling process. 
\item $msg$ : A message, $msg$,  is simply a sequence of $blk\_sz$ number of bytes. It is read as an array of type $\tau$. The capacity of the array depends up on the storage requirements of the element from $\tau$ i.e the capacity $C_{\tau} = blk\_sz/\mathcal S (\tau)$. Following operations are defined on msg array:
\begin{itemize}
\item size(msg) : number of elements stored in the message
\item full(msg) : check if $size(msg) == C_{\tau}$
\item add(msg) : add an element to the message $msg$, if there is room
\end{itemize}

\item $channel$ : Channel identifies a single session of communications between sender and receiver. Usually, a session between sender and receiver consists of transferring a complete stream by breaking it into a sequence of messages and transferring those one by one. Each message exchange between sender and receiver is tagged with the channel identifier to uniquely qualify this session.

\item $send(msg, reciever, channel)$ : A blocking call that sends the bits stored in msg to receiver. The communication between the sender and receiver is identified by the channel variable. The receiver should post a receive command (shown next) with the same channel number and sender field set to either $ANY$ i.e. receive from any sender, or to the specific sender in order to receive the message from this sending box. 
  
\item $msg \longleftarrow recv(sender, channel)$ : The corresponding receive blocking call to receive message in this channel. In general, there can multiple simultaneously active channels between the same sender, receiver pair.

\item $reader$ : A reader is a simple encapsulation over the receive operation. We need this concept to contrast the buffered reader approach, presented in section~\ref{sec:buffered_reader}, with the one without it. The encapsulation allows us to keep the design and algorithmic description of build CSR unchanged. Following operations are associated with $reader$ object:
\begin{itemize}
\item $reader \longleftarrow init(channel)$ : initialize a reader to read messages on a particular  $channel$
\item $read(reader, sender)$ : read data from $sender$ on the channel with which the reader is initialized with.
\end{itemize}

%recieve over the MPI
\end{itemize}

\subsection{A note of presentation of abstractions and algorithms }
Before we proceed to a detailed description we briefly visit the presentation style followed in the rest of the paper. Essentially, we have opted for a functional programming style for description of all our concepts and algorithms as opposed to the more frequently used imperative style of programming. 
The functional style is a better fit as it naturally  allows us to express generic (type agnostic) operations and  pipelining. Both these are crucial for our work.  Furthermore, it makes the presentation of algorithms terse and highlights core computation by reducing the amount of unnecessary syntactic expressions.

However, in order to not let the functional practices come in the way of presentation we make use of functional expressions in a restricted setting.
To begin with, a {\em generic type} is represented using symbol $\tau$. 
All the functors have arity at most three. Moreover, we only use two functional programming specific constructs, viz., bind and apply. 
Bind takes a functor  and an object (which can either be another functor or an object instance) and returns a new  functor of lesser arity by binding the corresponding argument to the instance. This operation is denoted using symbol $\bind$. 
For example, $sort(cmp, iter)$ takes two arguments: one, a comparison function and, the other, an iterator. The expression $\bind(sort, int\_cmp, \_1)$ returns a new  sorting function 
that has arity one, i.e, it only takes  $iter$ as an argument. The sorting (of the collection represented by $iter$) by this new function is dictated by the $int\_cmp$ function. 
The function $apply(F, C)$ (denoted by $\apply$) takes a single arity functor $F$ and an iterator. It returns a new collection whose elements are derived from applying $F$ to elements in the collection.

A few additional terms used here are as follows:

\paragraph*{tuples}
A  tuple, as usual, represents an ordered sequence of values. The function
$first, second, third, \dots$ extract the first, second, third and so on field of the tuple. 

\paragraph*{iterators} 
The collection of elements of basic data types or of tuples  is represented using iterators. The interface of iterators consists of:
\begin{itemize}
\item $init$ : Takes a collection or stream  and initializes it to corresponding iterator (integer or edges) 
\item $eos$ : Tells if the iterator has reached the end
\item $next$ : advance the iterator to next element (int or edge)
\item $get$ : returns the current element or object
\item $xget$ : Is derived from the get function. In returns tuple $(iter, get(iter))$. 
\end{itemize}

Any operator, using these functions, can access every single element of the collection represented using the iter.  For example to simply scan the elements the following code can be used:

\begin{tabbing}
\=XXXX\=XXXX\=XXXX\= \kill
$\func{scan}(iter)$:\\
1. $while(!eos(iter))$\\
2. \> \> $x \longleftarrow get(iter)$\\
3. \> \> $\text{do some work using x}$\\
4. \> \> $next(iter)$\\
\end{tabbing}

\paragraph*{Functors}
Following functors are used in this paper
\begin{itemize}
\item $filter(p, iter)$ : $p$ is a predicate and $iter$ is an iterator representing a collection. $filter$ represents a new iterator consisting of elements from the collection for which the predicate holds true.
\item $enum(iter)$ : enumerates the elements present in the iter collection. 
\item $uniq(iter)$ : Assuming iter represents a sorted sequence of elements, unique creates a new iterator where the duplicates are removed from the input sequence. 
\item $seq_{nb}$: An iterator that represents sequence of integer $[0:nb-1]$.
\end{itemize}

\subsection{Streams and Iterators}
We now begin with definitions of streams and their iterators   designed specifically to address the 
problem of building scalable CSR representation on SSD-enabled cluster of compute nodes such as Gordon~\cite{Norman:2010:ADS:1838574.1838588}.

\paragraph*{(Byte) Stream}
The most basic,  non-trivial, type is  a  (byte) stream object. A stream object 
is a sequence of bytes and can be of two types:
\begin{itemize}
\item Transient : A transient stream arrives over the network and can be read only once.
\item Persistent : A persistent stream is stored on physical media and can be read several times.
\end{itemize}
A transient (or in-network) stream is identified by $(source, destination, channel\_id)$. A persistent stream is addressed using attributes set $(file\_name, sz, off)$. Variable $file\_name$ is the name of the file that store the stream which is of size $sz$ bytes and starts at offset $off$.
We define following operations on a stream :
\begin{itemize}
\item store : Store a persistent or transient stream to a new location on the physical media
\item load  : load the stream into the main memory (assuming sufficient memory)
\item split : split a stream into several small streams of a fixed given size
\end{itemize}

We describe several iterators and discuss their implementation in our framework.  Each iterators also encapsulates an operation; for example, the
$in\_network$ iterator encapsulates the operation of transferring a persistent 
stream from one box to another. 
Similarly, the $sort\_merge\_join$ iterator encapsulates a join operation over
two streams. The instantiation of an iterator gives the program a handle to its  execution. However, instantiation does not execute the operation. The execution of 
the operation happens when a  scan is performed on the iterator. 
In this respect the iterators can seen as  non-blocking operators. Moreover, they can be pipelined 
which in our case turn out to be crucial for overcoming main memory constraints.

\paragraph*{Formatted Iterators}
Formatted iterators, such as integer iterator (int\_iter) or edge iterator (edge\_iter),  defines iterator 
object over the byte stream. 
A specialization of formatted iterators are the random access iterators. Random access iterators support two additional 
functions: size and get\_at. Size returns the number of elements in the iterator while get\_at returns the element at
the required index. Random access iterators provide the vector/array functionality. In order to derive 
a random access iterator over the stream it is necessary (for performance reasons) to load the stream in memory.

\paragraph{em\_stream\_iter (iter\_esi)}
This iterator is used to scan elements of type $\tau$ stored as stream on external memory. In this work, $\tau$  can be integer or edges. The state of the iterator is defined by   variables $cursor, curr\_blk, ptr, sz$. The entire stream is broken into blocks. Variable $curr\_blk$ points to currently active block.  A block is made active by memory mapping its content.  $ptr$ points to  this memory mapped region of curr\_blk portion of stream and read as an array of elements of type $\tau$.  Variable  $cursor$ indexes into the array starting at $ptr$ while $sz$ represents the total size of the stream in bytes. 

The working of iterator functions $init, eos, get,$ and $clean$ are evident from the code description. 
The function $next$ advances the cursor. If it is the end of the block, then the current block is unmapped and the next block is mapped into the memory. The cursor is reset back is zero (line 2--6).

\begin{tabbing}
\=XXXX\=XXXX\=XXXX\= \kill
$\func{init}(fn, sz, off)$:\\
1. \> \> $iter\_esi \longleftarrow \lbrace cursor=0,$ \\
   \> \> $curr\_blk=0,ptr= mmap(fn,offset,blk\_sz) \rbrace$\\
$\func{next}(iter_{esi})$:\\
1. $cursor = cursor + 1$\\
2. $if(cursor * \mathcal S(\tau) == blk\_sz)$\\
3. \> \> $munmap(ptr, blk\_sz)$\\
4. \> \> $curr\_blk = curr\_blk + 1$\\
5. \> \> $ptr = mmap(fn, curr\_blk * blk\_sz, blk\_sz)$\\
6. \> \> $cursor = 0$\\
$\func{eos}(iter_{esi})$:\\
1. $if(curr\_blk* blk\_sz + cursor * \mathcal S(\tau) < sz)$\\
2.\> \> $return\ false$\\
3. $return\ true$\\
$\func{get}(iter\_esi)$:\\
1. $return\ ptr[cursor]$\\
$\func{clean}(iter\_esi)$:\\
1. $munmap(ptr,sz)$
\end{tabbing}

\newcommand{\zip}{\zeta}

\paragraph{Sorted merge iterator}
Sorted merge iterators represent a sorted sequence of elements obtained from merging an  input  collection of iterators (of the same type). It assumes that the input iterators are also represents  sorted sequences.  A scan operation over this iterator performs a sorted merge of the input sequence. %\TBD{iterators as sequence}

The set $\lbrace H, in\_iters \rbrace$  encapsulates the state of this iterator. 
$H$ represents a heap object while $in\_iters$ is a  collection of  iterators over which the sorted merge operation is performed. The heap $H$ maintains a collection of $(iter, value)$ tuples, where the $value$ is the current value accessed from the iterator, i.e., $value = get(iter)$. 
 We assume following operations on the heap object:
\begin{itemize}
\item $top$ returns the top element of the heap
\item  $pop$: removes the top elements and returns its copy
\item  $insert$: insert a $(key, value)$ pair in the heap
\end{itemize}
The algorithm implemented here is the canonical sorted-merge, i.e., retrieve 
the element from the stream that has smallest elements among all the current elements of the stream. We refer to a standard textbook for further details~\cite{Silberschatz:2005:DSC:993519}. 

\begin{tabbing}
\=XXXX\=XXXX\=XXXX\= \kill
$\func{init}(in\_iters)$:\\
1. \> \> $iter_{smi} \longleftarrow \lbrace H, in\_iters \rbrace$\\
2. \> \> $\apply(\bind(insert, H, \_1), \apply(xget, \apply( $\\ 
\> \> \> $\bind(filter, eos, \_1), in\_iters)))$\\
$\func{next}(iter_{smi})$:\\
1. $if(!eos(first((top(H)))$\\
2. \> \> $insert(H, xget(next(first(pop(H)))))$\\
3. \> \> $return$\\
4. $pop(H)$\\
$\func{eos}(iter_{smi})$:\\
1. $if(empty(H))$\\
2.\> \> $return\ true$\\
3. $return\ false$\\
$\func{get}(iter_{smi})$:\\
1. $return\ second(top(H))$\\
$\func{clean}(iter_{smi})$:\\
1. $while(!eos(iter_{smi})$\\
2. \> \> $next(iter_{smi})$\\
\end{tabbing}

\paragraph{in-network stream iter}
\label{sec:in_network_iter}
In-network stream iterators are used to scan sequences of elements sent by another thread running within the same or remote box. 
The set $\lbrace  sender, cursor, msg, reader(channel) \rbrace$  encapsulate the state of this iterator. The  variable $channel$ identifies this communication, i.e., sender and receiver issue send and receive command  parametrized with the  $channel$  value.

\begin{tabbing}
\=XXXX\=XXXX\=XXXX\= \kill
$\func{init}(channel, sender)$:\\
1. \> \> $iter_{nsi} \longleftarrow \lbrace reader(channel),$ \\
\> \> $cursor=0,msg = buf(blk\_sz) \rbrace$\\
2. \> \> $msg = read(reader, sender)$\\
$\func{next}(iter_{nsi})$:\\
1. $cursor= cursor+1$\\
2. $if(cursor == c_{\tau})$\\
3. \> \> $msg = read(reader,sender)$\\
4. \> \> $cursor = 0$\\
$\func{eos}(iter_{nsi})$:\\
1. $if(size(msg) < c_{\tau})$\\
2. \> \> $if(cursor == size(msg))$\\
3. \> \> \> $return\ true$\\
4. $return false$\\
$\func{get}(iter_{nsi})$:\\
1. $return\ second(top(H))$\\
$\func{clean}(iter_{nsi})$:\\
1. $while(!eos(iter_{nsi}))$\\
2. \> \> $next(iter_{nsi})$\\

\end{tabbing}

\paragraph{sort-merge-join iter}
The join operator is required to perform the relabeling of source and destination. In this paper, we adhere to a  restricted notion of join: 
a join between streams  $S$ and $J$ given functions $field_S, field_J,$ and $join_{SJ}$,   results in a new stream whose elements are derived as follows:  $Join(S,J)=$
$$\lbrace join_{SJ}(s,j)\ \thickvert\  field_S(s) == field_J(j), s \in S, j \in J \rbrace$$.

Furthermore, we assume that the streams are sorted based on the join attributes i.e elements in stream $S$ are ordered according to values in $field_S$ of each element. Similarly, for stream $J$. We now describe the
sort-merge-join implementation  that produces the resulting stream assuming 
this condition  holds.  The by-product of a sort-merge-join operation is that
the resulting stream is sorted based on join field (the $field_S, field_J$  for stream $S$ and $J$, or the field corresponding to these in new stream). %\TBD{what is Join field}.

The set $\lbrace in\_iter, out\_iter, join\_fn, in\_join\_field,$ $out\_join\_field\rbrace$  encapsulates the state of this iterator. $in\_iter$ and $out\_iter$ represents the inner and outer input (sorted) streams (or relations), i.e., streams $S$ and $J$. The function $join\_fn$ takes two tuples, one from each relation, to produce an output relation. In addition, comparison operators 
are required over the join fields. In our case it is the canonical integer or integer derived comparison operator.  The following algorithm describes the 
textbook implementation of sort-merge-join  using in 
our proposed framework. The two streams are read simultaneously. When the join criteria is satisfied the join\_fn is called with arguments set as the current elements of the streams.  The result of join\_fn  produces the current element for the resulting iterator.

\begin{tabbing}
\=XXXX\=XXXX\=XXXX\=XXXX\= \kill
$\func{init}(in\_iter, out\_iter, join\_fn, in\_join\_field, $\\
\> \> \> $out\_join\_field)$:\\
1.  $iter_{smji} \longleftarrow \lbrace in\_iter, out\_iter, join\_fn, in\_join\_field,$\\
\> \> $out\_join\_field \rbrace$\\
2. $if(!eos(out\_iter))$\\
3. \> \> $while(in\_join\_field(in\_iter) < $\\ 
\> \> \>  $out\_join\_field(out\_iter))$\\
4. \> \> \> $next(in\_iter)$\\
$\func{next}(iter_{smji})$:\\
1. $next(out\_iter)$ \\
2. $if(!eos(out\_iter))$\\
3. \> \> $if(out\_join\_field(get(out\_iter)) == $\\
\> \> \> \>  $in\_join\_field(get(in\_iter)))$\\
4. \> \> \>  $return$\\
5. \> \> $while(in\_join\_field(in\_iter) < $\\
\> \> \> $out\_join\_field(out\_iter))$\\
6. \> \> \> $next(in\_iter)$\\
$\func{eos}(iter_{smji})$:\\
1. $return\ eos(out\_iter)$\\
$\func{get}(iter_{smji})$:\\
1. $return\ join\_fn(get(in\_iter), get(out\_iter))$\\
$\func{clean}(iter_{smji})$:\\
1. $clean(in\_iter)$\\
2. $clean(out\_iter)$
\end{tabbing}
%\textcolor{red}{This is red text}
%\TBD{a brief explanation of the working of algorithm}. \TBD{the importance of clean}.

The generic implementation of sort-merge-join allows us to perform join over any kind of streams: in-network, external memory, sorted-merge, or, even a stream obtained from sort-merge-join operation. This generality and the iterator based implementation would be crucial towards developing efficient and pipelined algorithm for building CSR. We shall join an in-network stream (identifier map stream) with an external memory edge stream in order to relabel the edges.

\subsection{Blocking Operators}

Here, we define two communication operations over the streams. These operators  behave differently than the  iterator based operators where the instantiation is decoupled from execution,
and therefore allows for pipelining, i.e., several iterators can be instantiated at once and chained together so that all of them execute in pipelined fashion. 
The blocking  operators, on the other hand,  are executed upon call and the calling process is blocked 
until the operation is complete.

\paragraph*{broadcast\_stream}
The broadcast operation takes as input a stream and sends  a  copy 
to every other box. In order to do so the stream is partitioned into blk\_sz messages. Each message is broadcasted over the network. The ordering of messages is maintained due to the use of blocking compunction operation, viz. MPI\_Send and MPI\_Recv, from the runtime.

\begin{tabbing}
\=XXXX\=XXXX\=XXXX\=XXXX\=XXXX\= \kill
$\func{broadcast\_stream}(iter)$:\\
1. $msg = buf(blk\_sz)$ \\ 
2. $while(!eos(iter))$ \\
3. \> \> $add(msg, get(iter))$ \\
4. \> \> $if(full(msg))$ \\
5. \> \> \> $\apply(\bind(MPI\_Send, msg,\_1), seq_{nb})$ \\
6. \> \> \> $reset(msg)$ \\
7. \> \> $next(iter)$ \\
8. $\apply(\bind(MPI\_Send, msg,\_1), seq_{nb})$ \\
\end{tabbing}

\paragraph*{scatter\_stream}
The scatter stream takes an additional mapping function which maps elements of the stream to a unique box. The input stream is partitioned into $nb$ sub streams based upon the mapping function. The sub streams are send to their respective boxes.

\begin{tabbing}
\=XXXX\=XXXX\=XXXX\=XXXX\= \kill
$\func{scatter\_stream}(iter, map\_fn,channel\_id)$:\\
1. $M \longleftarrow \lbrace msg_1, \dots, msg_{nb} \rbrace$ where $msg_{i} = buf(blk\_sz)$ \\
2. $while(!eos(iter))$ \\
3. \> \> $o \longleftarrow map\_fn(get(iter))$ \\
3. \> \> $add(msg_{o}, get(iter))$ \\
4. \> \> $if(full(msg_{o}))$  \\
5. \> \> \> $MPI\_Send(msg_{o}, o, channel\_id)$ \\
6. \> \> \> $reset(msg_{o})$\\
6. \> \> $next(iter)$ \\
7.  $\apply(\apply(bind(MPI\_Send, \_1, \_2, channel\_id), seq_{nb}), M)$\\
8.  $\apply(reset, M)$\\

\end{tabbing}
These operators require a thread at the receiving box to collect the incoming 
messages via the MPI\_Recv primitive. This is achieved by invoking a new thread which  scans a in-network stream iterator (described  earlier in section~\ref{sec:in_network_iter}) whose state variables (sender and channel) set accordingly i.e. sender is set to the rank of the box broadcasting (or scattering) the stream and channel is set to a value that uniquely identifies this communication.

\section{Edge list to Distributed CSR : The algorithms}
We now describe the approach for building the distributed CSR representation using the proposed iterators and operators.
The algorithm ties the various iterators and operators in a way that allows for pipelined 
distributed processing and shared memory/parallel processing. 

Our approach consists of five phases: setup, assign unique local identifiers,
relabel edges (destination), relabel edges (source), and finally build CSR. 

\subsubsection{The driver routine}
The buildCSR procedure  is invoked on the console with requisite parameters such as number of boxes, number of cores per box, the channel identifier 
for input edgelist. As with any MPI-based program, the runtime launches a process on each of the box with the same set of parameters. Each box is also
assigned a unique id from the range $[0:nb]$ which is known as the rank of the process. 

Each MPI process executes the phases, in that order, and in sync with other processes. 
It may also launch $nc$ pthreads either for parallel processing or for pipelined processing.

\subsubsection{Setup}
This step collects the edges generated by external processes and sets all the system parameters accordingly. 
The external processes generates the edges as a transient stream.
This stream is split into $nc$ stream, one per core worker. The core worker transforms the received stream into a persistent 
stream by storing in a physical location. 

\subsubsection{Assign unique local identifier}
In this phase each label is mapped to a unique box where it is a assigned a  unique numeric identifier.  In order to do so all the labels mapped to a box are maintained at that box in sorted order. 

The steps to distribute the labels across boxes and  build this sorted ordering is as follows:
Each box  partitions the edge stream (maintained in external memory) into   sub streams of size $mmc$ bytes. 
Each of these sub streams are converted into integer random access iterator by loading the sub stream
into the main memory and then casting into a integer iterator. The stream is then sorted in the memory and then written back to the external storage in a new location. The computational expression being:

\begin{tabbing}
\=XXXX\=XXXX\=XXXX\=XXXX\= \kill
1. $op = \bind(save, \bind(sort, \bind(int\_iter, \bind(load, \_1))))$ \\
2. $S1 \longleftarrow \apply(op, split(S, mmc))$
\end{tabbing}
where $S$ is the input edge stream on the box. 
%The expression  $\apply(\bind(store,\bind(\bind(sort, cmp), \bind(iter_{int}, load))),S)$ captures the steps carried out during the sort identifier step.

Next we scatter the identifiers based upon a mapping function. The scattering is performed in a manner such that they arrive at the destination in sorted order. To achieve this,  each box performs a sorted merge over streams in $S1$ and then calls the blocking scatter operation:

\begin{tabbing}
\=XXXX\=XXXX\=XXXX\=XXXX\= \kill
1. $ scatter(sorted\_merge(S1),$\\
\> \> $ LABEL\_SCATTER\_CHANNEL)$\\
\end{tabbing}

The scatter operation at a box produces a  sub-stream for every other box. Hence each box $b$ receives $nb$ streams, one from each scatter operation at a box (see figure~\ref{fig:comm_pattern}).
 These streams contain labels (in sorted order) that are mapped to box $b$.
A collector thread, at each box,  opens $nb$ in-network stream iterator, to receive the in-coming streams.  The collector also  performs a  sorted merge over these streams to arrive at a single stream of sorted labels. The pseudo-code describe the procedure for the collector thread:

\begin{tabbing}
\=XXXX\=XXXX\=XXXX\=XXXX\= \kill
1. $R1 \longleftarrow sorted\_merge(\apply(\bind(in\_network\_stream$\\
\> \> $(LABEL\_SCATTER\_CHANNEL, \_1)), seq_{bn}))$ \\
\end{tabbing}

Assigning unique identifier from 
range $[0,t]$, where $t$ is the total number of unique labels in the resulting stream can be done efficiently. Specifically, the computational expression 
is 
$enumerate(uniq(R1))$. 
We observe that with initializing $R1$ we have only so far instantiated a handle on the entire step and at that we haven't yet executed it. 
A scan over $R1$ i.e.  $scan(R1)$  will execute the step and produce a sorted stream of labels on the box. 

We call the stream $R1$ as {\em identifier map} as it contains the mapping from labels to new numeric identifiers.

\subsubsection{Relabel edges (destination field)}
We now proceed to relabel the destination with the new identifier i.e.
the numeric identifiers for the labels determined in the previous step.
If $(u,v)$ is an labeled-edge s.t. $v$ is assigned (hashed) to box $b$ and has numeric identifier $nid$ then, in this step, we relabel this edges as $(u,<b,nid>)$. 

We achieve this by performing a  join over two streams, $E$ and $I$, where $E$ represents an stream of edges ordered by destination and $I$ represents a stream of {\em all} (identifier, label) tuple sorted by labels. The result of this  join operation is a new stream of edges with destination relabeled to new identifier i.e

\begin{multline}
join(E,I) = \lbrace (u, \langle bid, nid \rangle ) \thickvert (u,v) \in E,\\ (v, \langle bin,nid \rangle) \in I \rbrace
\end{multline}

Since both the streams are sorted on join field we can obtain $join(E,I)$ by perform sort-merge-join over the streams $E$ and $I$ i.e. 

\begin{multline}
join(E,I) 
= sort\_merge\_join(E, I, relabel\_des,\\ second, first)
\end{multline} 
where 
relabel\_des is a function with following signature: 
\begin{tabbing}
\=XXXX\=XXXX\=XXXX\=XXXX\= \kill
$\func{relabel\_des}(e, <l,id>)$:\\
1. $return\ (e.src, id)$ if $e.des == l$\\
\end{tabbing}

The algorithm below describe the procedure of building stream $E$. 
The local edgelist is partitioned into a collection of streams each of size $mmc$ (represented by function $chunk\_partition$).  This collection is further divided among the participating cores. Each core then sorts the streams assigned to it and materializes at a new location. A sorted merge over these streams results in $E$ i.e. a stream of edges sorted by the destination field. 

\begin{tabbing}
\=XXXX\=XXXX\=XXXX\=XXXX\= \kill
1. $C \longleftarrow chunk\_partition(split(S, mmc), nc)$\\
2. $op = \bind(save, \bind(\bind(sort, des\_cmp, \_1), $\\
\> \> \> \> $\bind(edge\_iter, \bind(load, \_1))))$ \\
3. $\text{for each thread, tid, where } tid \in [0:nc]$\\
4. \> \> $S2 \longleftarrow \apply(op, C[tid])$\\
5. $\text{wait for all threads}$\\
6. $E \longleftarrow sorted\_merge(S2)$\\ 
\end{tabbing}

We observe that stream $I$ can be obtain by performing sorted merge over 
the  identifier maps derived in the previous ``assigned unique identifier'' phase. However, we cannot perform 
sorted merge directly over  the streams since they are distributed across the boxes. 
In order to bring all the identifier map stream together (on every box) each box broadcasts its identifier map stream to all other boxes.
Boxes also launch another thread to receive the identifier maps. It then performs a sorted merge over these streams to arrive at stream $I$. The computational expression to derive $I$ is 

\begin{tabbing}
\=XXXX\=XXXX\=XXXX\=XXXX\= \kill
1. $I \longleftarrow sorted\_merge(\apply( \bind(in\_network\_stream($\\
\> \> $ {\textrm {\tiny IDMAP\_BCAST\_CHANNEL}}, \_1), seq_{nb})))$
\end{tabbing}

Hence the computation  of this step on each box is distributed across two threads: first thread receives the sorted labels and performs sorted merge to build identifier map. It also broadcast the identifier maps. The second thread performs the sort-merge join over stream $E$ and $I$ which in turn receives all the identifier maps. 
Figure 2 illustrate this execution on a box. The first thread performs sorted merge and broadcasts the identifier map (inside boxes shown in top right corner). The second thread's operation is illustrated in detail. The sorted-merge edge stream (from below) and sorted-merge identifier map stream (from right) are joined 
together to produce a relabeled edge. 

\begin{figure}
\includegraphics[width=3.0in]{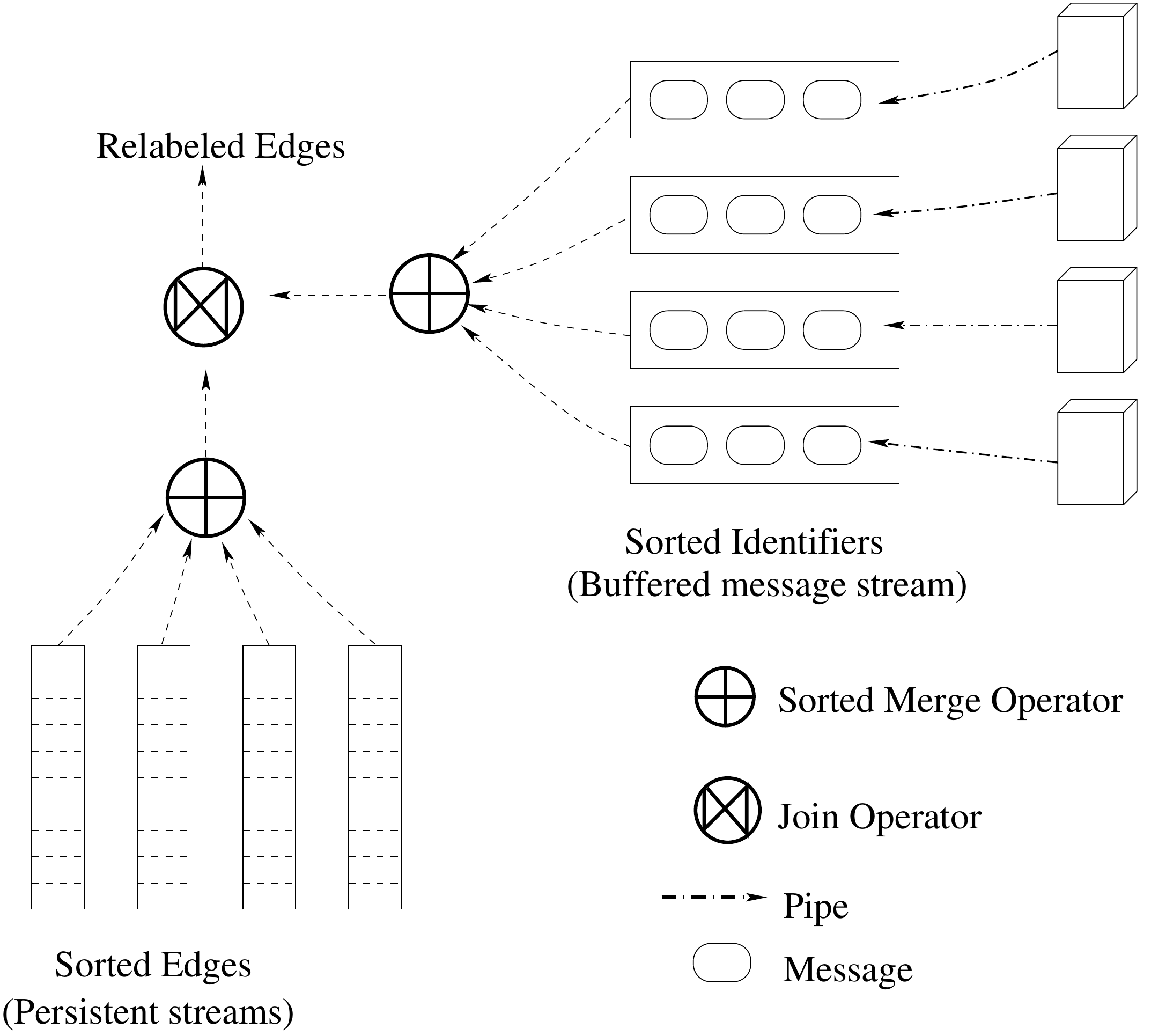}
\caption{Illustration of relabeling performed by each core worker. Identifier stream is broadcasted by each box which is
received by the cores using buffered pipes. The core worker performs sorted merge over the in-network stream 
and over the persistent edge stream. The two sorted streams are joined using sort-merge-join style algorithm to produce the relabeled stream.}
\end{figure}

\subsubsection{Relabel edges (source field)}
We now relabel the source field  of the edges. This step is similar to the step of relabeling destination with minor differences: sort the edges (this time based on source field), launch thread to broadcast identifier maps, perform sort merge join over two streams: One obtained by sorted merge of edge stream and another obtained by sorted merge of identifier stream. As opposed to previous step, in this step we only derive the handle for sort\_merge\_join operations, i.e., derive the iterator but do not perform the scan operation over this iterator. The scan operation is performed in the next scatter edge phase.

\subsection{Scatter relabeled edge and build CSR}

In this step, we redistribute the relabeled edges obtained in the previous step so that each edge is stored at its owner box. 
Additionally, the edges received at a box during the scatter operation are sorted based upon the new identifiers. This is the because the new identifiers are assigned in the ascending order of the labels and that we use sort-merge-join algorithm for relabeling. 
This expedites building the CSR representation.

Each box performs  
$scatter(S)$ operation.  This operation breaks the relabeled edge (source field) stream into sub-streams, one for each compute box. Edges arriving at a box are owned by that box. Furthermore, as explained earlier they are also sorted based upon the new identifier. 
Hence, in order to derive the desired edge stream on each box we only need to additionally perform a sorted merge over the incoming edge stream i.e. in a separate thread  we execute the following operation:
\begin{multline}
R \longleftarrow sorted\_merge(\apply( \bind(in\_network\_stream,\\
 {\textrm{\tiny EDGE\_SCATTER\_CHANNEL}}, \_1), seq_{bn}))
\end{multline}

\subsection{Avoiding communication deadlocks}
\label{sec:buffered_reader}
The above description although algorithmically correct does have the problem of reaching a deadlock state, i.e, threads may form a circular dependency of waits 
using the blocking MPI communication primitives MPI\_Send and MPI\_Recv.

\begin{SCfigure}
\centering
  \includegraphics[width=1.50in]{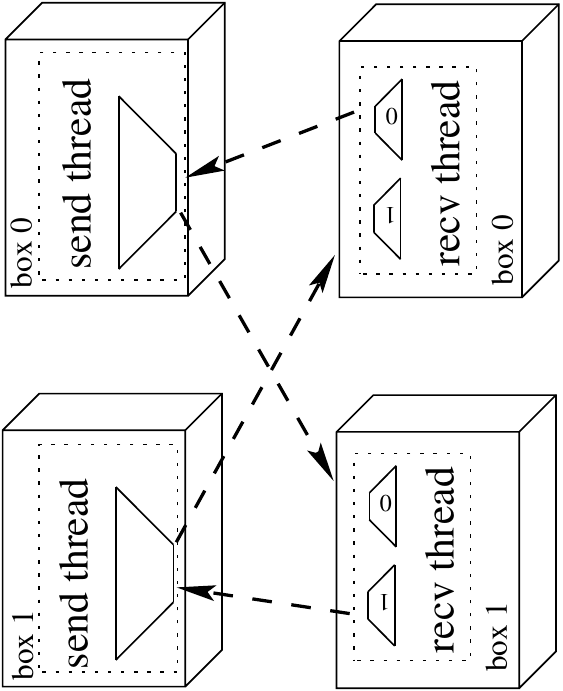}
  \caption{Illustration of deadlock scenario in our system. Two processes each with two threads, one for scattering and one for receiving. The receiver thread has two (MPI message) readers one for sender 0 and one of sender 1.  The dotted lines show the dependency from originating thread to the waiting thread, for example receive thread on box 0 is waiting for send (scatter thread) on box 1. Hence 
dependency lines goes from box 0 receive thread to box 1 send thread. \label{fig:deadlock}}

\end{SCfigure}

We show this through a simple setup consisting of two boxes each having two threads one for sending data and the other for receiving. All the 
threads are communicating over the same channel. 
Figure~\ref{fig:deadlock} illustrates this dependency cycle. The arcs of this dependency cycle are:
\begin{enumerate}
\item Receiver on box 1 is blocked after it has issued an $MPI\_Recv(sender = 1)$ command,  i.e, it is waiting for sender on box 1 to respond. Hence, we draw an arc from 
the receiver (on box 1) to the sender (on box 1). 

\item The sender on box 1 is blocked after it has issued an $MPI\_Send(reciever = 0)$ command, i.e., it is waiting for receiver on box 0 to receive the message. Hence, we have an arc from sender on box 1 to receiver on  box 0.

\item The receiver on box 0 has posted a $MPI\_Recv(sender = 0)$ and therefore blocked. Hence, we have an arc from receiver on box 0 to sender on box 0.

\item The sender on box 0 is waiting for reviver on box 1 in order to complete the communication call $MPI\_Send(receive = 1)$. Hence, there is an arc from 
sender on box 0 to receiver on box 1. 
\end{enumerate}
These dependencies together  represent a deadlock situation where the algorithm makes no further progress. 

We present an enhancement to our   scheme to break the circular dependency without changing its architecture and methodology. 
We introduce a  buffered reader object that reads messages on behalf of in\_network streams. 
This reader is shared by all the in\_network streams on a box having the same channel. 

Below we show the complete implementation of the buffered reader.
The state of buffered reader is 
determined by variables $channel, msg\_queues, msg\_free\_pool,$ and  $msg\_allocated$.  
 The reader serves the request to read the next message from stream originating from sender $s$ as follows:  it maintains a first in first out (FIFO) queue of messages per sender. If the queue for sender $s$ is not empty then it simply pops a message from the sender's queue and returns it. However, if the queue is empty then then the reader issues  repeated $MPI\_Recv(sender=ANY)$ calls until it receives a message from the sender $s$. This message is returned to the caller. 
The rest of the received messages are placed in the queues corresponding to its sender.

\begin{tabbing}
\=XXXX\=XXXX\=XXXX\= \kill
$\func{init}(channel, msg\_queues, msg\_free\_pool, msg\_allocated  )$:\\
1. \> \> $reader \longleftarrow \lbrace channel, msg\_queues, $\\
\> \> $msg\_free\_pool, msg\_allocated \rbrace $ \\
2. $msg\_allocated[seq_{nb}]  = NULL$\\
$\func{read\_msg(reader,sender)}$:\\
1. $if(msg\_allocated[sender])$\\
2. \> \> $push(pool, msg\_allocated[sender])$\\
3. \> \> $msg\_allocated[sender] = NULL$\\
4. $if(!msg\_queues[sender].empty())$\\
5. \> \> $msg \longleftarrow msg\_queue[sender].pop()$\\
6. \> \> $msg\_allocated[sender] = msg$\\
7. \> \> $return\ msg$\\
8. $while(1)$\\
9. \> \> $if(!pool.empty())$\\
10.\> \> \> $msg = pool.pop()$\\
11. \> \> $else$\\
12. \> \> $msg = init()$\\
13. \> \>  $source \longleftarrow MPI\_Recv(msg, $\\
\> \> \> $MPI\_ANY\_SOURCE, channel)$\\
14. \> \> $if(source  == sender) break$\\
15. \> \>  $msg\_queues[source].push(msg)$\\
16. \> \> $return msg$\\
\end{tabbing}

\section{Operation Pipeline in MPI/Pthread Primitive}
Pipelined processing is an important optimization principle in any data processing system. Our choice  of data types (streams and its variants), operators (sorted-merge, sort-merge-join, broadcast, and, scatter) and  the algorithms for various phases of building CSR are designed to make pipeline processing feasible. We study our proposed approach as a data  workflow and show how  MPI multi-threaded interface can be  used towards pipelined processing without incurring any additional synchronization overheads.

The entire computation for building CSR on a box can be decomposed into stages. 
A stage is defined by a group of threads distributed across the boxes  performing the same step of computation. All the stages/threads are active simultaneously and communicate with each other through the data distribution operator (scatter and broadcast) and buffered reader operator on different channels.
 
Figure~\ref{fig:pipeline} shows the various stages  and the channels over which the distribution and reader operators communicate in our proposed approach.
The pipeline stages (in that order) are merge and redistributed labels, build-and-broadcast idmap, relabel and scatter edges, and, merge-and-build-csr. The threads in the first and second stages communicate over the {\tiny{LABEL\_SCATTER\_CHANNEL}}, threads in second and third stage communicate over {\tiny{IDMAP\_BCAST\_CHANNEL}} and those in third and last stage communicate over {\tiny{EDGE\_SCATTER\_CHANNEL}}. 

\begin{SCfigure}
\centering
\includegraphics[width=1.50in, height=2.000in]{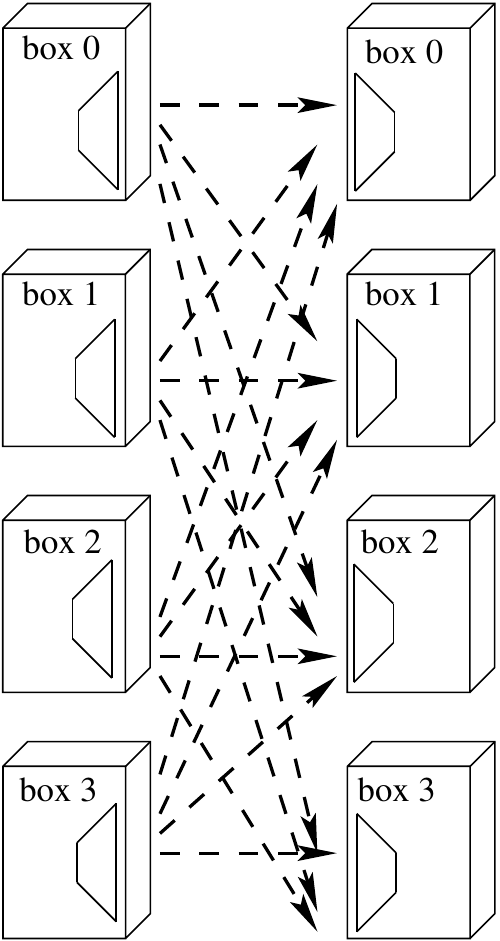}
\caption{The communication pattern in our framework is a complete bipartite graph. The sender threads across boxes send messages to every  receiver threads on the channel.\label{fig:comm_pattern} }
\end{SCfigure}

All the data-distribution threads interfacing a channel communicate with all the readers plugged to the channel. Hence the communication pattern at any channel is a complete bipartite graph $K_{nb,nb}$ as shown in figure~\ref{fig:comm_pattern}.

A general method for pipelining operations emerges from this discussion. If the operation can be decomposed into sequence of stages  such that the input iterator of a stage comes from the output iterator of the previous stage then it can be implemented in the MPI/Pthread framework using the distributed operators and buffered readers such that it allows for pipelined distributed computing.

The added advantage of the pipelining using our scheme is that it eliminates the need for synchronization between threads executing the stages. Pipelining in multi-threaded environment invariably require synchronization since the data is produced (output as result) by one thread and is consumed (as input) by another thread. Hence, these two threads are in producer consumer relationship and need to synchronize the access to the concerned data~\cite{Graefe:1990:EPV:93597.98720}.
In our approach the threads performing each stage of the computation do not need perform any explicit synchronization.  All synchronization needs are handled transparently through the MPI multi-threaded environment and the blocking send/recv calls.

\section{Experiments}
We conduct experiments over the proposed edgelist to CSR converter to study the impact of 
pipelining and overheads due to shared memory and distributed computing on its performance. In addition,
we also compare our algorithm with that of the standard distributed edgelist  to CSR routine 
provided in the parallel boost graph library (PBGL).  PBGL is distributed graph library 
build on top the Boost framework which is a collection of data structures, algorithms, system utilities, and 
advanced language primitives. 
Since PBGL is a well used library and has an (unoptimized) standard implementation of edgelist to CSR implementation
we consider it to be a good reference point for comparison.

All experiments are conducted on Gordon supercomputer~\cite{Norman:2010:ADS:1838574.1838588}
Each compute node contains two 8-core 2.6 GHz Intel EM64T Xeon E5 (Sandy Bridge) processors and 64 GB of DDR3-1333 memory. 
The compute nodes mount a single 300 GB SSD (280 GB usable space). The latency to the SSDs is several orders of magnitude lower than that for spinning disk ($<100$ microseconds vs. milliseconds).

In our experiments we used synthetic random and scale-free graph edgelist generator.
We also varied various algorithm parameters such as  scale, edge factor, block size, $mmc$, and the number of boxes.
The algorithm description so far had assumed one MPI process per box. However, it is possible to launch several MPI process on a single box.
We compare performance in both the  settings:  multiple MPI process on a single box vs. multiple boxes each with one MPI process.

\begin{figure}
  \includegraphics[width=2.00in,angle=-90]{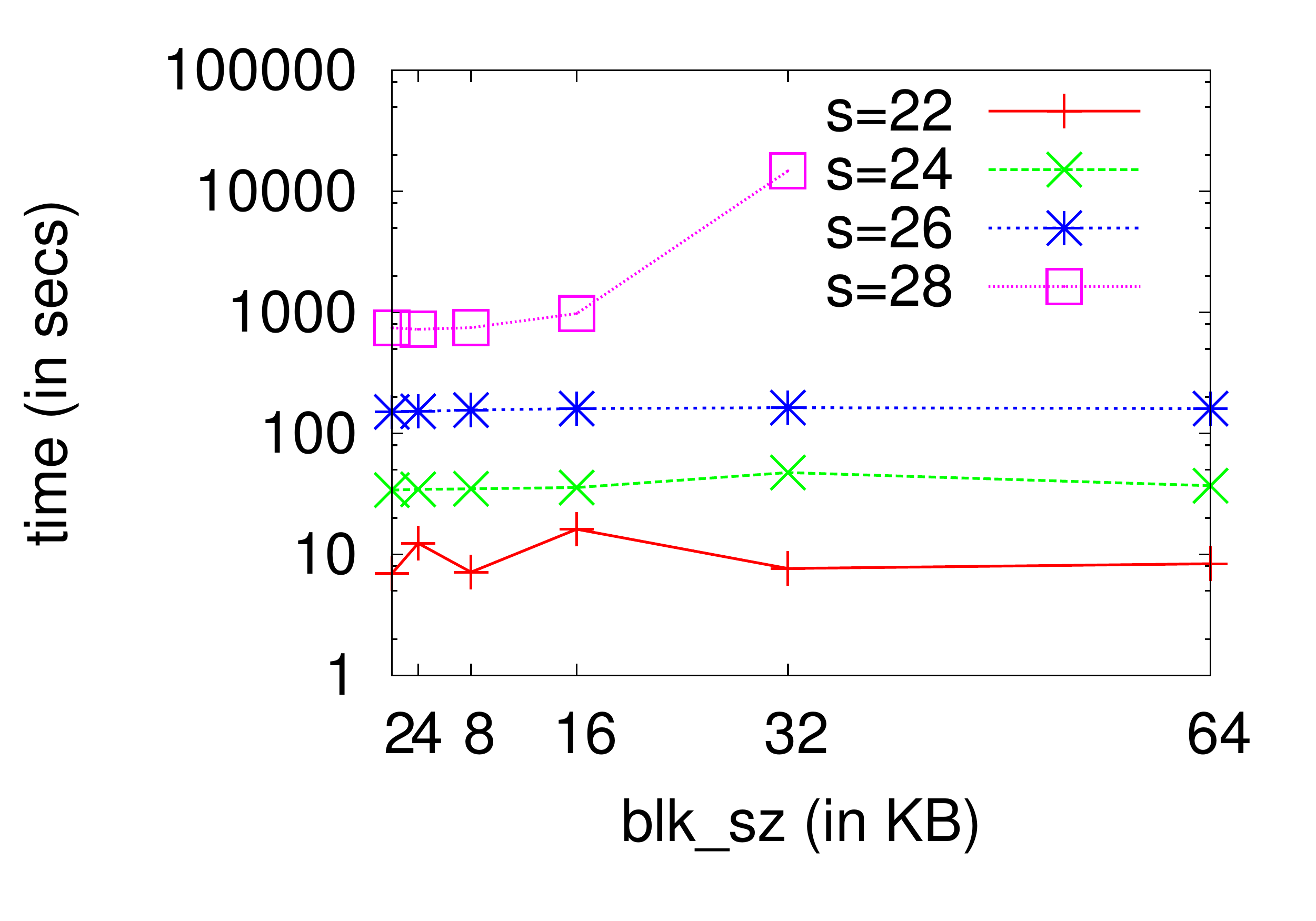}
  \caption{Build CSR time using one mpi process vs blk\_sz for various scale graph\label{fig:single_box_blk_box}}
\end{figure}

In all the experiments we use the random edge generator unless otherwise specified. The edge factor is set to 8. 
Figure~\ref{fig:single_box_blk_box} shows performance for a single mpi process for various scale edge list varying the block size. 
The edge factor is fixed to 8. We see that the system can gracefully handle a scale $28$ edge list. The total storage cost of this graph 32 GB. 
The total communication cost is determined by both the size of edge list and the blk\_sz. At larger block size fewer messages are exchange but 
each message exchange is of longer duration. We see that the system favors smaller blk\_sz and that at large blk\_sz the system 
performance deteriorates considerably.

The reason why higher blk\_sz is disadvantageous in our  algorithm  lies in the  implementation of the hybrid MPI/pthread
scheme. Currently, this scheme is implemented using global lock which serializes all communication, i.e, only one MPI call
can be active at any given instance. When the blk\_sz is large a single MPI call continues for a long duration. The
rest of threads, after finishing their local work, block and wait for this MPI thread to complete. This leads to a scenario where
the rest of the threads stall for longer period of time and therefore suffer decreased performance. This aspect hinders the performance and scalability of
our system considerably. However, we conjecture that the current implementation of hybrid MPI/pthread is sub-optimal and in 
future implementations in which  a single  active MPI call does not block rest of the calls the algorithm will scale with
respect to large blk\_sz as well.

\begin{figure}
\centering
\subfigure[Single box \label{fig:single_box_vs_mpip}]{
  \includegraphics[width=1.60in,angle=-90]{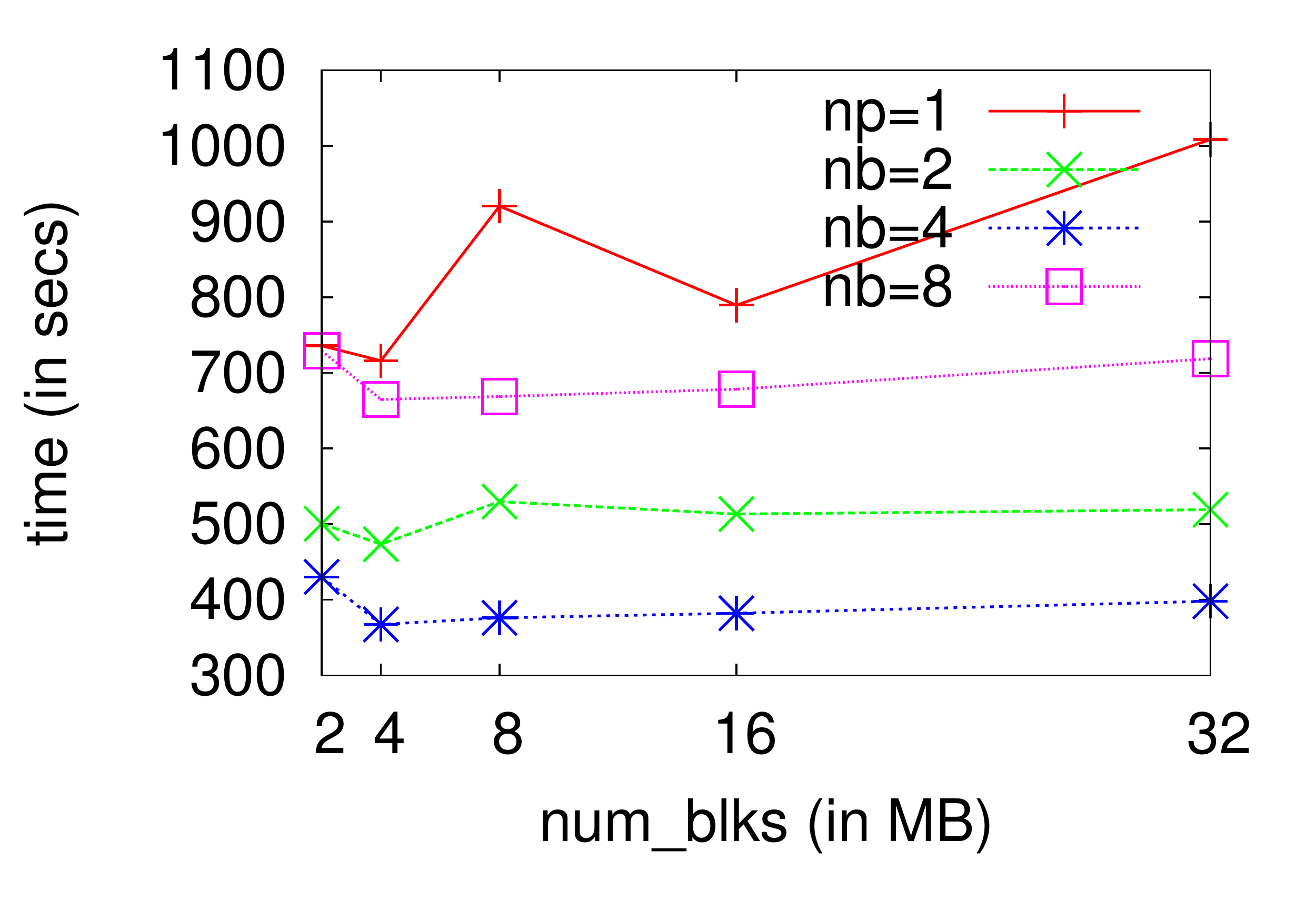}
}
\subfigure[Multiple box \label{fig:multi_box}]{
  \includegraphics[width=1.60in,angle=-90]{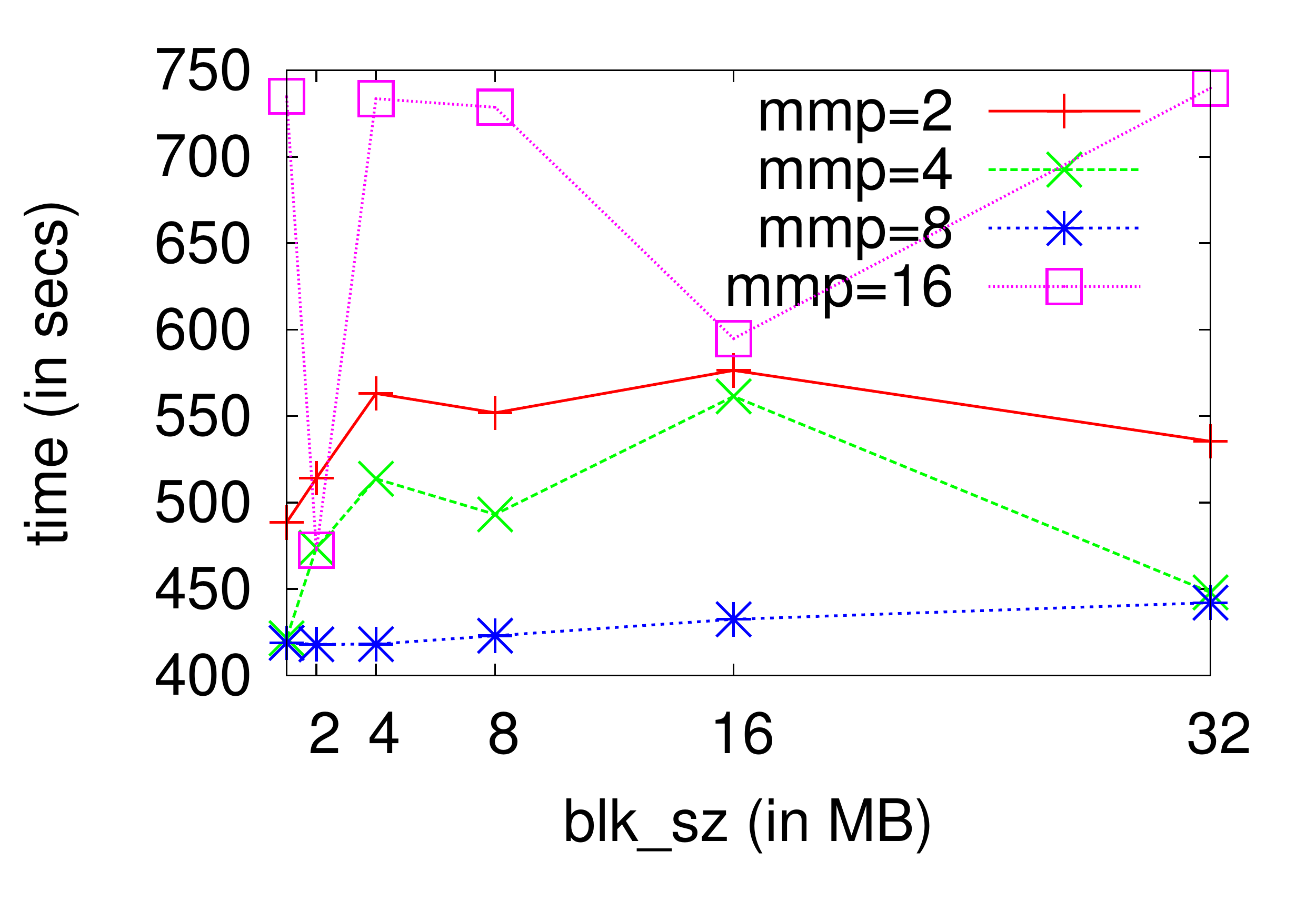}
}

\caption{Build CSR time on a single box and multiple boxes varying  the number of mpi processes. }
\end{figure}

In figure~\ref{fig:single_box_vs_mpip} we study the strong scaling aspect of the algorithm on a single box, i.e, we build CSR representation for scale $28$ with varying number of processes on a single box. 
The algorithm achieves scalability until 4 MPI process and performance drops when the number of processes is increased to 8. This is because at 8 MPI process amounts to 32 active threads on a single node. In this configuration we have over subscribed the number of threads since  the compute node only as 16 cores and  thus reducing the performance. 

We also conduct experiment  varying the number of boxes for a fixed scale graph. In this setup, there is one MPI process per box.  We observe a small improvement with two boxes. 
 As mentioned earlier, due to high overhead associated with communication in MPI/pthread runtime we do not see strong scaling across the number of boxes. Our hope is that with future improvements to MPI/pthread  better scalability is possible.

Although due to communication costs our scheme currently does not scale beyond two nodes or two MPI processes, we show that it has
good scalability with respect to scale parameter when compared with the standard algorithm present in PBGL. First, we note that
irrespective of the number of nodes used the boost graph algorithm cannot scale beyond scale 26  graph. On the other hand our scheme 
can generate scale $30$ graph on single node and even larger using multiple nodes if communication overheads can be tolerated. We see that 
PBGL's time graph quadratically with the graph size due to the use sort function on the entire edge list and non-pipelined implementation. 
Our scheme on the other hand shows much better scalability with respect to increase in scale. We attribute this primarily to 
sort-merge-join and pipelined implementation. 

\begin{figure}
\mbox{
\subfigure[num\_box=1]{
\includegraphics[width=1.15in,angle=-90]{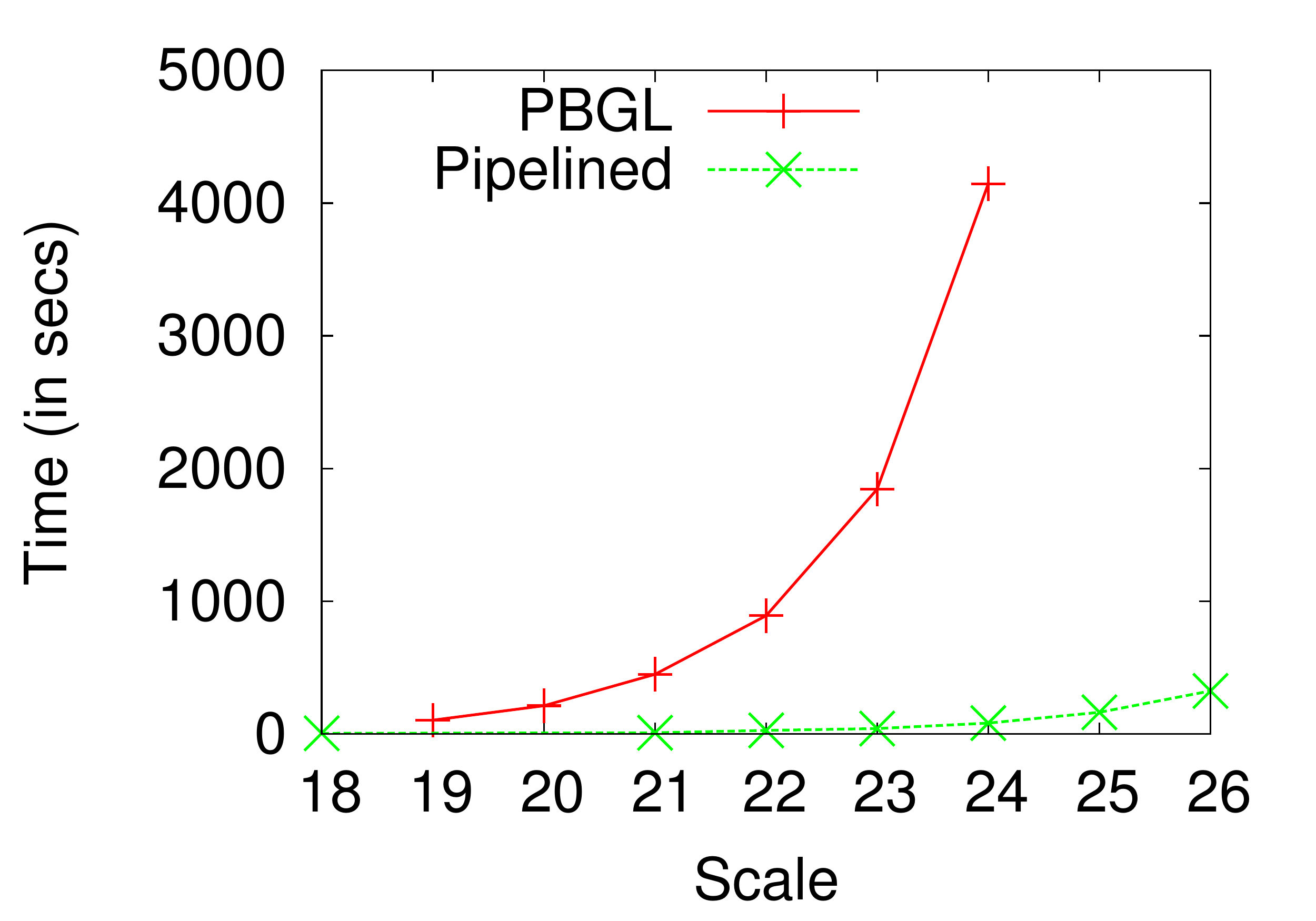}
}
\subfigure[num\_box=2]{
\includegraphics[width=1.15in,angle=-90]{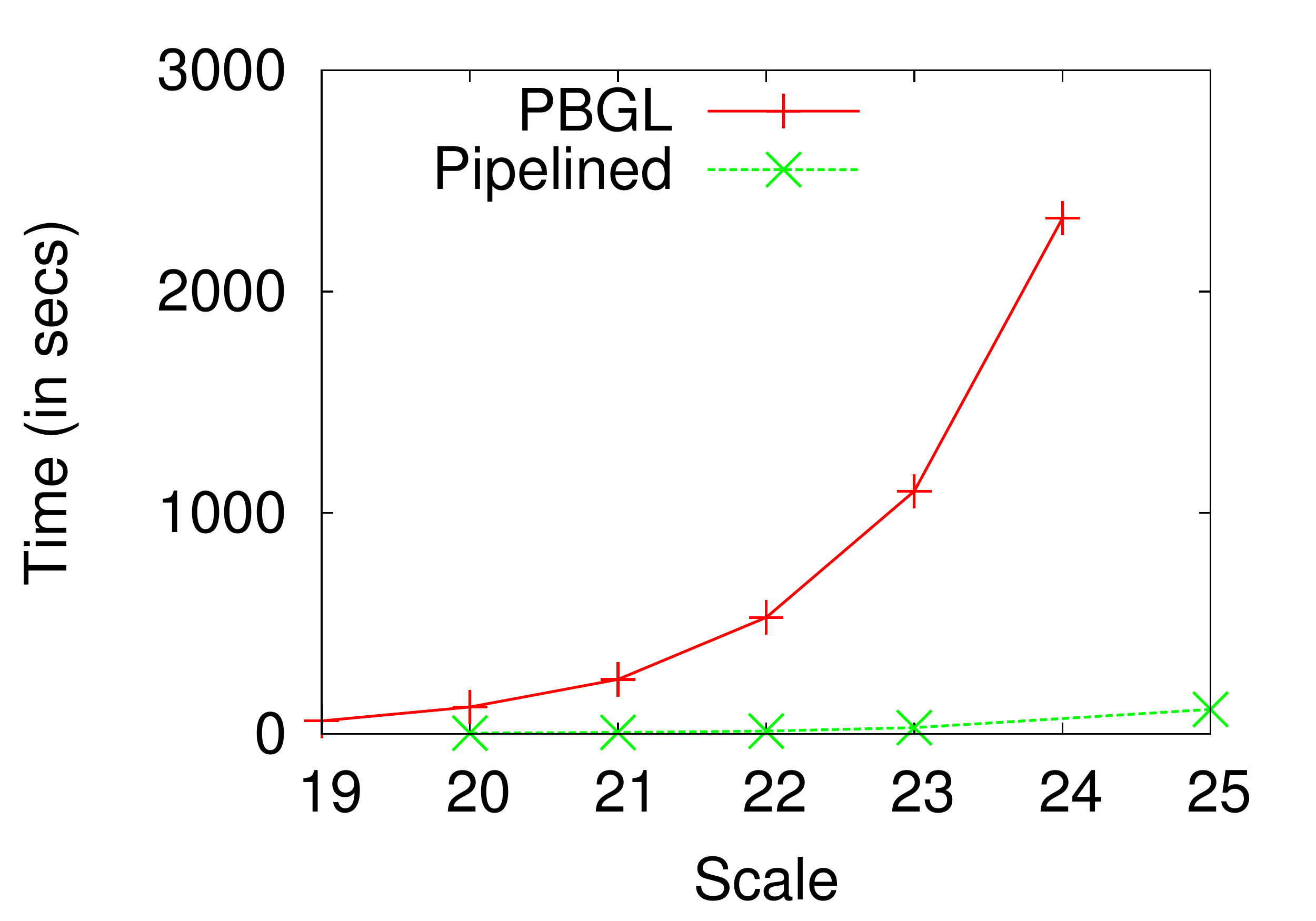}
}
}
\caption{Comparison with PBGL's distributed RMAT generator routine using one and two mpi process. Both schemes perform edge generation and sorting. Our scheme additionally 
performs relabeling and read and write to SSDs. Permutation is turned off for PBGL's RMAT generator for fair comparison. Additionally, PBGL cannot generate graph large than scale 25 (assuming 8 bytes identifier) while our scheme can generate graph up to scale $30$.}
\end{figure}

\begin{SCfigure}
  \includegraphics[width=1.50in,angle=-90]{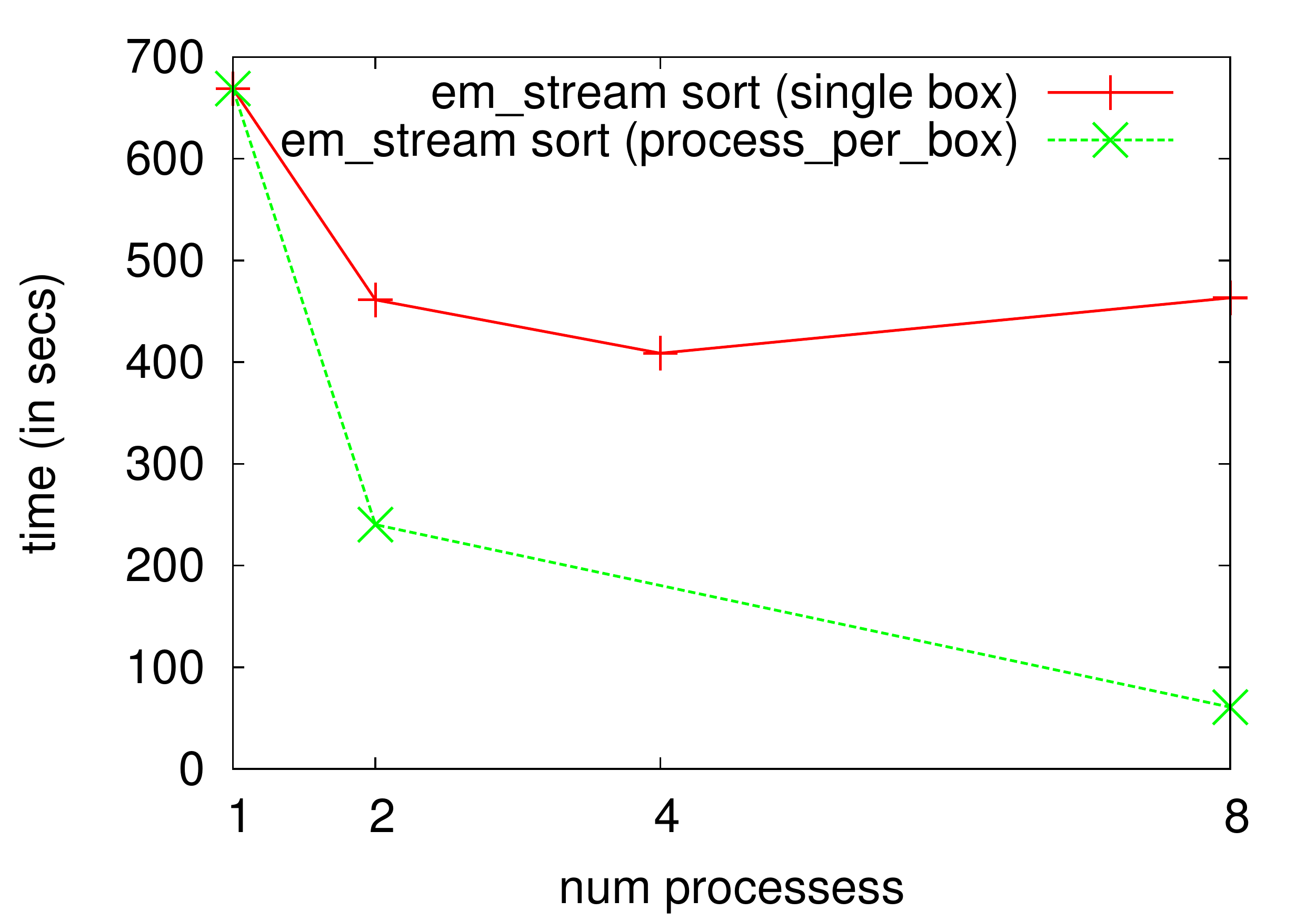}
  \caption{Sort edges and identifier stream time on single box (multiple process) and multiple box (one proces per box)\label{fig:sort}}
\end{SCfigure}

Figure~\ref{fig:sort} compares performance for the first phase of the buildCSR routine, i.e., the sort labels and sort edges, which do not require
communication across MPI process. It compares the performance for multiple MPI process on single box vs. multiple boxes, one MPI process per node. 
As we increase the number of processes on a single node due to limited resources (cores, bandwidth on SSDs etc) the performance flattens out. 
However, as there is proportionally more resources on latter setup we see increased scalability.

Figure~\ref{fig:timestamp} illustrates the pipeline processing of the algorithm. We ran a single MPI processing for scale $20$ edge list.
In this run every send and receive message is an event that is depicted as upper or lower triangle respectively. 
Color denote the channel of communication. 
We see that the messages across channel are interleaved together implying a smooth pipelined processing.

\section{Conclusion and Future Work}
We have shown a pipelined algorithm for constructing distributed graph representation from a given collection of edge list. 
Our implementation is based up on the hybrid MPI/pthread runtime that allows for multiple threads to simultaneously invoke
communication calls. Through experiments we demonstrate good scalability and 4--6 times speedup
on single and two compute nodes when compared with standard PBGL implementation. We highlight the deficiency of the current implementation
of the MPI/pthread runtime which hinders scalability beyond two compute nodes. 
We are currently investigating the use of asynchronous I/O and other runtimes to bypass this limitation.
Our scheme has implication on general data-intensive computing that we can currently investigating. 

\bibliographystyle{abbrv}
\bibliography{putting_together}

\begin{thebibliography}{10}

\bibitem{Buluc:2011:CBD:2076556.2076566}
A.~Bulu\c{c} and J.~R. Gilbert.
\newblock The combinatorial blas: design, implementation, and applications.
\newblock {\em Int. J. High Perform. Comput. Appl.}, 25(4):496--509, Nov. 2011.

\bibitem{Calvanese:2009:ODD:1611703.1611710}
D.~Calvanese, G.~Giacomo, D.~Lembo, M.~Lenzerini, A.~Poggi, M.~Rodriguez-Muro,
  and R.~Rosati.
\newblock Reasoning web. semantic technologies for information systems.
\newblock chapter Ontologies and Databases: The DL-Lite Approach, pages
  255--356. Springer-Verlag, Berlin, Heidelberg, 2009.

\bibitem{sparql_graph_homomorphism}
O.~Corby and C.~Faron-Zucker.
\newblock Implementation of sparql query language based on graph homomorphism.
\newblock In U.~Priss, S.~Polovina, and R.~Hill, editors, {\em Conceptual
  Structures: Knowledge Architectures for Smart Applications}, volume 4604 of
  {\em Lecture Notes in Computer Science}, pages 472--475. Springer Berlin /
  Heidelberg, 2007.

\bibitem{Dolby:2008:SGC:1483155.1483189}
J.~Dolby, A.~Fokoue, A.~Kalyanpur, L.~Ma, E.~Schonberg, K.~Srinivas, and
  X.~Sun.
\newblock Scalable grounded conjunctive query evaluation over large and
  expressive knowledge bases.
\newblock In {\em Proceedings of the 7th International Conference on The
  Semantic Web}, ISWC '08, pages 403--418, Berlin, Heidelberg, 2008.
  Springer-Verlag.

\bibitem{Graefe:1990:EPV:93597.98720}
G.~Graefe.
\newblock Encapsulation of parallelism in the volcano query processing system.
\newblock In {\em Proceedings of the 1990 ACM SIGMOD international conference
  on Management of data}, SIGMOD '90, pages 102--111, New York, NY, USA, 1990.
  ACM.

\bibitem{gregor06:pbgl_siampp_presentation}
D.~Gregor and A.~Lumsdaine.
\newblock The parallel boost graph library.
\newblock Presentation at the SIAM Conference on Parallel Processing, San
  Francisco, California, February 2006.

\bibitem{scalable-subgraphs}
J.~Huang, D.~J. Abadi, and K.~Ren.
\newblock Scalable sparql querying of large rdf graphs.
\newblock PVLDB, 4(21), August 2011.

\bibitem{triple/Store/Report}
R.~Lee.
\newblock Scalability report on triple store applications.
\newblock Technical report, Massachusetts Institute of Technology, 2004.

\bibitem{Low:2012:DGF:2212351.2212354}
Y.~Low, D.~Bickson, J.~Gonzalez, C.~Guestrin, A.~Kyrola, and J.~M. Hellerstein.
\newblock Distributed graphlab: a framework for machine learning and data
  mining in the cloud.
\newblock {\em Proc. VLDB Endow.}, 5(8):716--727, Apr. 2012.

\bibitem{Madduri11}
K.~Madduri.
\newblock Snap (small-world network analysis and partitioning) framework.
\newblock In D.~A. Padua, editor, {\em Encyclopedia of Parallel Computing},
  pages 1832--1837. Springer, 2011.

\bibitem{Malewicz:2010:PSL:1807167.1807184}
G.~Malewicz, M.~H. Austern, A.~J. Bik, J.~C. Dehnert, I.~Horn, N.~Leiser, and
  G.~Czajkowski.
\newblock Pregel: a system for large-scale graph processing.
\newblock In {\em Proceedings of the 2010 ACM SIGMOD International Conference
  on Management of data}, SIGMOD '10, pages 135--146, New York, NY, USA, 2010.
  ACM.

\bibitem{sparql}
S.~Malik, A.~Goel, and S.~Maniktala.
\newblock A comparative study of various variants of sparql in semantic web.
\newblock In {\em Computer Information Systems and Industrial Management
  Applications (CISIM), 2010 International Conference on}, pages 471 --474,
  oct. 2010.

\bibitem{DBLP:conf/semweb/McBride01}
B.~McBride.
\newblock Jena: Implementing the rdf model and syntax specification.
\newblock In {\em SemWeb}, 2001.

\bibitem{Neumann:2008:RRE:1453856.1453927}
T.~Neumann and G.~Weikum.
\newblock Rdf-3x: a risc-style engine for rdf.
\newblock {\em Proc. VLDB Endow.}, 1(1):647--659, Aug. 2008.

\bibitem{Norman:2010:ADS:1838574.1838588}
M.~L. Norman and A.~Snavely.
\newblock Accelerating data-intensive science with gordon and dash.
\newblock In {\em Proceedings of the 2010 TeraGrid Conference}, TG '10, pages
  14:1--14:7, New York, NY, USA, 2010. ACM.

\bibitem{Quilitz:2008:QDR:1789394.1789443}
B.~Quilitz and U.~Leser.
\newblock Querying distributed rdf data sources with sparql.
\newblock In {\em Proceedings of the 5th European semantic web conference on
  The semantic web: research and applications}, ESWC'08, pages 524--538,
  Berlin, Heidelberg, 2008. Springer-Verlag.

\bibitem{Rohloff:2007:ETT:1780453.1780500}
K.~Rohloff, M.~Dean, I.~Emmons, D.~Ryder, and J.~Sumner.
\newblock An evaluation of triple-store technologies for large data stores.
\newblock In {\em Proceedings of the 2007 OTM Confederated international
  conference on On the move to meaningful internet systems - Volume Part II},
  OTM'07, pages 1105--1114, Berlin, Heidelberg, 2007. Springer-Verlag.

\bibitem{Rohloff:2010:HMS:1940747.1940751}
K.~Rohloff and R.~E. Schantz.
\newblock High-performance, massively scalable distributed systems using the
  mapreduce software framework: the shard triple-store.
\newblock In {\em Programming Support Innovations for Emerging Distributed
  Applications}, PSI EtA '10, pages 4:1--4:5, New York, NY, USA, 2010. ACM.

\bibitem{OntDB}
N.~Sarda.
\newblock Ontology-enabled database management systems.
\newblock In R.~Sharman, R.~Kishore, and R.~Ramesh, editors, {\em Ontologies},
  volume~14 of {\em Integrated Series in Information Systems}, pages 563--584.
  Springer US, 2007.

\bibitem{Silberschatz:2005:DSC:993519}
A.~Silberschatz, H.~Korth, and S.~Sudarshan.
\newblock {\em Database Systems Concepts}.
\newblock McGraw-Hill, Inc., New York, NY, USA, 5 edition, 2006.

\bibitem{Stuckenschmidt:2005:TDP:1358613.1358616}
H.~Stuckenschmidt, R.~Vdovjak, J.~Broekstra, and G.~Houben.
\newblock Towards distributed processing of rdf path queries.
\newblock {\em Int. J. Web Eng. Technol.}, 2(2/3):207--230, Dec. 2005.

\bibitem{OCZ}
O.~technology.
\newblock {The Z-Drive R4}.
\newblock \url{http://www.ocztechnology.com/aboutocz/press/2011/445}, 2011.

\bibitem{Weiss:2008:HSI:1453856.1453965}
C.~Weiss, P.~Karras, and A.~Bernstein.
\newblock Hexastore: sextuple indexing for semantic web data management.
\newblock {\em Proc. VLDB Endow.}, 1(1):1008--1019, Aug. 2008.

\bibitem{Wood05kowari:a}
D.~Wood.
\newblock Kowari: A platform for semantic web storage and analysis.
\newblock In {\em In XTech 2005 Conference}, pages 05--0402, 2005.

\end{thebibliography}
\end{document}